\documentclass[showpacs,preprint]{revtex4}

\usepackage{graphicx}
\usepackage{amssymb}
\def\be{\begin{equation}}
\def\ee{\end{equation}}
\def\beq{\begin{eqnarray}}
\def\eeq{\end{eqnarray}}

\begin{document}


\title{Universality of isothermal fluid spheres in Lovelock gravity}


\author{Naresh Dadhich}  \email[]{nkd@iucaa.ernet.in}
\affiliation{ Jamia Millia Islamia University, Delhi, India \\ Inter-University Centre for Astronomy and Astrophysics, Pune, India \\
 Astrophysics and Cosmology Research Unit, 
 School of Mathematics, Statistics and Computer Science, University of KwaZulu-Natal,  Private Bag X54001, Durban 4000, South Africa}

\author{Sudan Hansraj} \email[]{hansrajs@ukzn.ac.za}
\affiliation{ Astrophysics and Cosmology Research Unit, School of Mathematics, Statistics and Computer Science,
University of KwaZulu-Natal, Private Bag X54001, Durban 4000, South Africa}

\author{Sunil D. Maharaj} \email[]{maharaj@ukzn.ac.za}
\affiliation{ Astrophysics and Cosmology Research Unit, School of Mathematics, Statistics and Computer Science,
University of KwaZulu-Natal, Private Bag X54001, Durban 4000, South Africa}


\date{\today}

\begin{abstract}
We show universality of isothermal fluid spheres in pure Lovelock gravity where the  equation of motion has only one $N$th order term coming from the corresponding Lovelock polynomial action of degree $N$. Isothermality is characterized by the equation of
state, $p = \alpha \rho$ and the property, $\rho \sim 1/r^{2N}$.  Then the solution  describing isothermal spheres, which exist only for the pure Lovelock equation, is of the same form for the general Lovelock degree $N$ in all dimenions $d \geq 2N+2$. We further prove that the necessary and sufficient condition for the isothermal sphere is that its metric is conformal to 
the massless global monopole or the solid angle deficit metric, and this feature is also universal.
\end{abstract}


\maketitle


\section{ Introduction}

Isothermal structures are supposed to model equilibrium and dynamically relaxed state of more complicated systems like stars, galaxies, cluster of galaxies and the universe as a whole. It has a long history in the works of several authors modeling these  structures as well as dark matter \cite{emden, plummer, saslaw, zwicky, sdm}, who have attempted to analyse
 different astrophysical and cosmological phenomena. It is an equilibrium  limiting state characterized by constant gravitational potential  (constant temperature). It translates into conditions with equation of state $p = \alpha \rho$ and $\rho  \sim 1/r^2 $ in Einstein gravity, and the physical configuration is a static sphere. Clearly it is not a fully physically satisfactory distribution as the density diverges, however the total mass contained inside a radius $r$ goes to zero at the centre. It is therefore a milder kind of singularity. The important point is that it captures the essential physical features of the limiting equilibrium state of  astrophysical and cosmological systems after a long enough time. The similar situation is the case with the uniform density sphere as a model of a star. This is also not entirely blemish-free  as the velocity of sound becomes infinite because of density being constant. Yet it beautifully models a star interior. \\

One of the most interesting generalizations of Einstein gravity is the Lovelock gravity \cite{lov} involving homogeneous polynomials of degree $N$ in Riemann curvature in action, yet remarkably the resulting equation of motion retains the desired second order character. No other generalization has this character, and hence Lovelock gravity is distinguished from all others. It however includes Einstein gravity in the linear order $N=1$ and quadratic Gauss-Bonnet (GB) for $ N=2$, and so on. Another remarkable feature of 
Lovelock gravity is that higher order terms are non-vacuous only in higher dimensions, $d \geq 2N+1$; i.e.  the Einstein gravitational equation is non-vacuous in $d \geq 3$ while GB in $d \geq 5$, and so on. A strong case has recently been made  \cite{nd, dgj} for the pure Lovelock equation being the right gravitational equation in higher dimensions. By pure Lovelock we mean that the equation has only one $N$th order term, there is no sum over for lower orders. Then it is envisaged that gravity has universal behaviour for all odd $d=2N+1$ and even $d=2N+2$ dimensions. For instance Einstein gravity is kinematic in $d=3$ dimension -- absence of non-trivial vacuum solution, this is in fact the case for all odd $d=2N+1$ dimensions relative to a properly defined $N$th order analogue of the Riemann curvature \cite{bianchi, cam-d, discern}. Further it turns out that several physical features of the Einstein gravity are carried forward in higher dimensions only by pure Lovelock gravity, for instance, existence of bound orbits around a static black hole \cite{bound} (also see \cite{probes}). \\

Lovelock gravity has been probed mostly in vacuum by several authors beginning with the E-GB black hole solution \cite{bd} and others \cite{whee, whit}, however there have also been some fluid solutions as well \cite{dkm, maha-hans, hans-maha, chil-hans}. In particular, universality of uniform density sphere was established  \cite{dkm}  for Einstein-Lovelock gravity by showing that it is always described by the  Schwarzschild interior solution. In more general terms pure Lovelock gravity is distinguished for overall gravitational behaviour being qualitatively the same in all odd and even $d=2N+1, 2N+2$ dimensions \cite{dgj, nd, discern}. It is therefore pertinent to study isothermal fluid distributions  for Lovelock gravity, and that is what we shall be concerned with in this paper. It turns out that for isothermal distributions  $\rho \sim 1/r^{2N}$  in all $d \geq 2N+2$ dimensions. This singles out, like existence of bound orbits \cite{bound}, pure Lovelock gravity. Further there can exist no isothermal sphere for the critical odd dimension $d=2N+1$; i.e. no isothermal distribution for $3$ and $5$ dimensions respectively for Einstein ($N=1$) and GB ($N=2$) gravity. Like the universality of Schwarzschild interior solution for uniform density sphere (which is universal in general Einstein-Lovelock gravity), we will establish universality of solution describing the  isothermal sphere for pure Lovelock gravity in all dimensions $d \geq 2N+2$. \\

The paper is organized as follows: In the next section we write the Lovelock equation for gravity followed by its specialization to the static fluid sphere. It readily follows that the critical odd
dimension $d=2N+1$ is not consistent with the isothermal properties, and then we show that the most general solution for isothermal sphere is the same for all  $d \geq 2N+2$ dimensions only in pure Lovelock gravity. Further we also show that the necessary and sufficient condition for isothermal sphere is that its metric is conformal to massless global monopole or solid angle deficit spacetime \cite{conform}, or gravitational potential $\lambda$ constant. We end with a discussion. \\


\section{ Lovelock gravity}

The Lovelock polynomial action \cite{lov} is given by the Lagrangian
\begin{equation}
\mathcal{L} = \sum ^N_{N=0} \alpha_N \mathcal{R}^{(N)}  \label{5}
\end{equation}
where
\begin{equation}
 \mathcal{R}^{(N)} = \frac{1}{2^N} \delta^{\mu_1 \nu_1 ...\mu_N \nu_N}_{\alpha_1 \beta_1 ... \alpha_N \beta_N} \Pi^N_{r=1} R^{\alpha_r \beta_r}_{\mu_r \nu_r}
\end{equation}
 and $R^{\alpha \beta}_{\mu \nu}$
 is the  $N$th order Lovelock analogue of the Riemann tensor as defined in Ref. \cite{bianchi}.  Also $ \delta^{\mu_1 \nu_1 ...\mu_N \nu_N}_{\alpha_1 \beta_1 ... \alpha_N \beta_N} = \frac{1}{N!} \delta^{\mu_1}_{\left[\alpha_1\right.} \delta^{\nu_1}_{\beta_1} ... \delta^{\mu_N}_{\alpha_N} \delta^{\nu_N}_{\left.\beta_N \right]}$ is the required Kronecker delta. \\

On variation of the action including the matter Lagrangian with respect to the metric,  it yields the equation of motion \cite{bianchi}
\begin{equation}
 \sum ^N_{N=0} \alpha_N \mathcal{G}^{(N)}_{AB} = \sum ^N_{N=0} \alpha_N \left( N\left(\mathcal{R}^{(N)}_{AB} - \frac{1}{2}\mathcal{R}^{(N)}g_{AB}\right)\right) =  T_{AB} \label{7}
\end{equation}
where $ \mathcal{R}^{(N)}_{AB} = g^{CD}\mathcal{R}^{(N)}_{ACBD}, \mathcal{R}^{(N)}=g^{AB}\mathcal{R}^{(N)}_{AB}$, and $T_{AB}$ is the energy momentum tensor. This is the gravitational equation in Lovelock gravity which corresponds to the cosmological constant for $N=0$, the Einstein equation for $N=1$ and to the Gauss-Bonnet equation for $N=2$, and so on.  \\

In particular the Einstein-Gauss-Bonnet equation is given by
\begin{equation}
G^{A}_{B} + \alpha H^{A}_{B} = T^{A}_{B}  \label{6}
\end{equation}
with metric signature $(- + + + ...)$ where $\mathcal{G}^{(2)}_{AB} = H_{AB}$,
\begin{equation}
H_{AB} = 2\left(R R_{AB} - 2R_{AC}R^C_B - 2R^{CD}R_{ACBD} + R^{CDE}_{A} R_{BCDE} \right) - \frac{1}{2} g_{AB} \mathcal{R}^{(2)}.
\end{equation}
In the next  section we shall employ the Lovelock equation for studying isothermal fluid sphere.

\section{Isothermal fluid sphere}

Isothermal spheres would be described by a static spherically symmetric metric which is given by
\be
ds^2 = - e^{\nu} dt^2 + e^{\lambda} dr^2 + r^2 d\Omega^2_{d-2} \label{8a1}
\ee
where $d\Omega^2_{d-2}$ is the metric on a unit $(d-2)$-sphere and  $\nu = \nu(r)$ and $\lambda = \lambda (r)$ are the metric potentials.  The energy momentum tensor for the
comoving fluid velocity vector $u^A=e^{-\nu/2} \delta^{A}_{0}$ has the form
$
T^A_{B}= \mbox{diag} \left( -\rho,\, p_r ,\,  p_{\theta} ,\,
p_{\phi},\, ...\right)$
for a neutral perfect fluid. Note that in view of spherical symmetry we have for all the $(d-2)$ angular coordinates $p_{\theta} = p_{\phi} = \cdots$. The conservation law   $T^{AB}_{}{}_{;B} = 0 $  gives  the  equation
\be
\frac{1}{2}\left( p_r + \rho\right)\nu' + p'_r + \frac{(d-2)}{r}\left(p_{r} - p_{\theta}\right) = 0 \label{8a11}
\ee
where a prime denotes  differentiation with respect to $r$. \\

We shall consider the pure Lovelock equation for a fixed $N$ without sum over lower orders. Gravity could couple with matter with only one coupling constant at one time, and hence we should have only one coupling constant for a fixed $N$ which means pure Lovelock gravity. There is also a strong case made for pure Lovelock gravity \cite{discern} by requiring universality of certain gravitational properties. In particular, gravity should be kinematic relative to $N$th order Lovelock-Riemann curvature in all critical odd $d=2N+1$ dimensions, this property singles out pure Lovelock gravity. We shall therefore concern ourselves only to pure Lovelock gravity with a fixed $N$ for the study of isothermal fluid spheres.  \\

For the above metric (\ref{8a1}), the pure Lovelock equation for a fixed $N$ from equation (\ref{7}) yields the expressions for density and radial pressure as
\begin{eqnarray}
\rho &=&\frac{(d-2) e^{-\lambda} \left(1-e^{-\lambda}\right)^{N-1}\left(rN\lambda' + (d-2N-1)(e^{\lambda} -1)\right)}{2r^{2N}} \label{9a} \\
p_r &=&\frac{(d-2) e^{-\lambda} \left(1-e^{-\lambda}\right)^{N-1}\left(rN\nu' - (d-2N-1)(e^{\lambda} -1)\right)}{2r^{2N}} \nonumber \\ \label{9b}
\end{eqnarray}
where we have set the corresponding coupling constant to unity. \\

It is clear that setting the potential $\lambda$ constant in (\ref{9a}) results in the density behaving  as $1/r^{2N}$. But is the converse also true? Let us rewrite the density expression as
\be
\rho =\frac{(d-2) [r^{(d-2N-1)}(1- e^{-\lambda})^N]'} {2r^{(d-2)}} = \frac{(d-2) [r^{(d-2N-1)}(1- e^{-\lambda})^N]'r^{2N+2-d}} {2r^{2N}}. \label{9c1}
\ee In order  for $\rho$ to go as $1/r^{2N}$ for isothermal behaviour, the numerator must be constant,
\be
(d-2)\left(r^{d-2N-1} \left(1-e^{-\lambda}\right)^N\right)' r^{2N+2-d} = K_1 \label{9c2}
\ee
and it integrates to give
\be
e^{-\lambda} = 1 -\left(\frac{K_1}{d-2N-2} + \frac{K_2}{r^{d-2N-1}}\right)^{1/N} \label{9c3}
\ee
where $K_2$ is an integration constant. Since $d\geq 2N+1$ always, and to avoid  a singularity at $r=0$, $K_2=0$. Hence the converse is also true. Therefore constant $\lambda$ is necessary and sufficient for isothermality. \\

We take $d\neq 2N+1$ and set $e^{\lambda} = k$, a constant, and write
\be
\rho = \frac{d-2}{2r^{2N}} (d-2N-1)(1 - 1/k)^N. \label{9l}
\ee
Since $p=\alpha\rho$, we write $p_r=\alpha\rho$, and solve for $\nu$ and then substitute it in the isotropy equation, $p_r=p_\theta$.  So we write by equating (\ref{9a}) and (\ref{9b}),
\be
\nu'=\frac{(k-1)(d-2N-1+\alpha)}{Nr} \label{9m}
\ee
which integrates to give
\be
e^\nu = r^\beta \label{9n}
\ee
where
\be
\beta=\frac{(k-1)(d-2n-1+\alpha)}{N} \label{9o}
\ee

Now from the conservation equation (\ref{8a11}) we obtain  $p_\theta$ as  given by
\beq
p_{\theta} &=& \frac{1}{4} r^{-2 N} \left(e^{\lambda}\right)^{-N} \left(e^{\lambda}-1\right)^{N-2} \left[-N r \lambda' \left\{2 (d-2 N-1) \left(e^{\lambda}-1\right)  \right. \right. \nonumber \\
&& \left. \left.
+r \nu' \left( e^{\lambda}-2 N +1\right)\right\} + \left(e^{\lambda}-1\right) \left\{-2 (d-2 N-1) (d-2 N-2) \left(e^{\lambda}-1\right) \right. \right. \nonumber \\
&& \left. \left.
+2N  ( d-2N-1) r\nu' +Nr^2 \nu'^2 +2N r^2 \nu''\right\}\right]
\eeq

Then the equation of pressure isotropy $p_r = p_{\theta}$ reads as
\beq
 r\lambda' \left\{2 (d-2 N-1)  \left(e^{\lambda}-1\right)+r\nu' \left( e^{\lambda}-2 N +1\right)\right\} && \nonumber \\
  -\left(e^{\lambda}-1\right) \left\{4 (d-2 N-1) \left(e^{\lambda}-1\right)+r \left(\nu' \left(-4 N+r \nu'+2\right)+2 r \nu''\right)\right\} &=& 0 \label{9d}
\eeq
\\

In view of the solution (13) and $\lambda$ being constant, the above equation determines the constant
\be
k= 1 + \frac{\beta^2-4N\beta}{4(d-2N-1)}.
\ee
Plugging in $\beta$ from equation (\ref{9o}), we ultimately obtain
\beq
k &=& 1 + \frac{N\beta}{d-2N-1+\alpha}, \nonumber \\
\beta &=& \frac{4\alpha N}{d-2N-1+\alpha}.
\eeq

Thus we have obtained the general solution for the pure Lovelock isothermal sphere. In particular for the critical even dimension, $d=2N+2$, these parameters take the form $k = 1 + N\beta/(1 + \alpha), \beta = 4\alpha N/(1 + \alpha)$. For the usual $4$-dimensional case of $N=1$, they take the same values as in Ref. \cite{sdm}; i.e. $k= 1 + \beta/{(1+\alpha)}$ and $\beta = 4\alpha/{(1+\alpha)}$.  \\

We have thus shown that the solution describing isothermal sphere has the same form in all $d\geq 2N+2$ dimensions establishing its universality for pure Lovelock gravity. It is also clear that for $d\neq 2N+1$, $\rho \sim 1/r^{2N}$ if and only if $\lambda$ is constant which means the necessary and sufficient condition for existence of isothermal sphere in pure Lovelock gravity is that $\lambda$ is constant. Furthermore the isothermality property of $\rho \sim 1/r^{2N}$ cannot be satisfied for different values of $N$ simultaneously, and hence the isothermal sphere can exist only for pure Lovelock gravity. \\

To complete the story, we now consider the case $d=2N+1$. \\

\subsection{ The case $d=2N+1$}

We wish to implement the isothermal perfect fluid conditions, $p_r=p_\theta=\alpha \rho \sim 1/r^{2N}$ for $d=2N+1$. From  (8) and (9), we get $\alpha \lambda' = \nu'$, and requiring $\rho \sim 1/r^{2N}$ leads to
\be
e^{-\lambda} = 1 - (k_1 \ln r + k_2)^{1/N}.
\ee
Substituting in the pressure isotropy equation (16) it turns out that it can only be satisfied for $k_1=0$. This means $\lambda$ constant leading to $\rho=0$ from  (8). Thus there can exist no isothermal sphere in the critical $d=2N+1$ dimension. This completes the proof that isothermal spheres exist only in dimensions $d\geq2N+2$, and the solution has universal validity. With $\lambda$ constant, it is solid angle deficit or massless global monopole spacetime \cite{dny} relative to Einstein gravity but it is flat relative to pure Lovelock gravity because the $N$th order Lovelock analogue of Riemann tensor vanishes  \cite{bianchi, cam-d, dgj}.  \\

\subsection{The case $\lambda$ constant}

We wish to probe the question: does there exist a fluid solution other than isothermal for constant $\lambda$? 
We can address this question by considering the isotropy equation (\ref{9d}) which for $e^{\lambda} = k$ reduces to
\be
2 r^2 \nu''  + r^2 \nu'^2  -2 (2N -1)r\nu' +4 (d-2 N-1) \left(k -1\right) = 0.
 \label{9f}
\ee
This is essentially a Riccati type equation and in this particular case admits the general solution
\be
e^{\nu} =   c_2 r^{ 2\left(N-\sqrt{A}\right)} \left(c_1+r^{2 \sqrt{A}}\right)^2  \label{101}
\ee
where $A = (d-2N-1)(1-k) +N^2$ and $c_1$ and $c_2$ are constants of integration. \\

From (\ref{9b}), the pressure is then given by
\be
p =\frac{(d-2) (k-1)^{N-1}  \left(c_1 \left(N-\sqrt{A}\right)^2+r^{2 \sqrt{A}} \left( N+\sqrt{A} \right)^2\right)}{2 \left(r^{2 \sqrt{A}}+c_1\right)k^{N} r^{2 N}} \label{102}
\ee
Clearly for it to describe an isothermal sphere, $c_1 = 0$ or $A=0$, and in either case it reduces to the isothermal solution (\ref{9l}). Could this describe a bound fluid sphere with its boundary given by $p=0$? That would require $c_1<0$ which then implies that pressure is singular at the center as well as another finite radius. This is no good. Thus with constant $\lambda$, isothermal sphere is the only possible solution for
$d\geq2N+2$ dimensions.

\section{ Conformal to massless global monopole}

It had been shown by one of us \cite{conform} that  the isothermal fluid sphere is always conformal to the spacetime of massless global monopole \cite{bv} or solid angle deficit \cite{dny} which also corresponds to constant potential in spherical symmetry; i.e.the metric potentials, $g_{tt}$ and $g_{rr}$ being constant. The global monopole  metric \cite{bv} is given by
\be
ds^2 = -f^2 dt^2 + \frac{dr^2}{f^2} + r^2 d\Omega^2, \, \, f^2 = 1 - \eta^2 - 2M/r
\ee
where $\eta$ is the global charge and $M$ is the mass of monopole. When $M=0$, the metric potentials become constant, then constant $g_{tt}$ could be absorbed by redefining $t$, while constant $g_{rr}$ produces non-zero curvature $R^{\theta\phi}{}{}_{\theta\phi} \sim 1/r^2$. It then corresponds to $\rho = -p_r \sim 1/r^2$, which is quite in line with isothermality except that one has to generate $p_{\theta} \sim 1/r^2$ and make $p_r>0$ and equal to $p_{\theta}$ through the conformal function. It turns out that this indeed happens. \\

Let us first  begin with the isothermal pure Lovelock sphere and then we show that it could be written conformal to the massless global monopole metric. It is of the form,
\be
ds^2 =  -r^{2b} dt^2 + k^2 dr^2 + r^2 d\Omega_{d-2}^2,
\ee
or equivalently,
\be
ds^2 = r^{2b} \left(-dt^2 + {\frac{k}{r^{2b}}}^2 dr ^2  + r^{2(1-b)} d\Omega_{d-2}^2 \right).
\ee
By defining $r^{1-b} = \bar r $, we can write
\be
ds^2 = {\bar r}^{2b/(1-b)} (-dt^2 + (\frac{k}{1-b})^2 d\bar r ^2  + \bar r^2 d\Omega_{d-2}^2)
\ee
Thus it is conformal to the massless global monopole metric. \\

Conversely we begin with a conformal metric
\be
ds^2 = f^2 (-dt^2 + k^2 dr ^2  +  r^2 d\Omega_{d-2}^2)
\ee
and define $r f=\bar r$, to write
\be
ds^2 = -r^{2b} dt^2 + \bar k^2 d\bar r ^2  +  \bar r^2 d\Omega_{d-2}^2
\ee
where $f = r^b$ and $b= k/\bar k -1$. This is the isothermal sphere metric. \\

We have thus shown that the necessary and sufficient condition for isothermal sphere is that its metric is conformal to the massless global monopole spacetime and this feature is universal for pure Lovelock gravity in all dimensions $d\geq2N+2$. It is interesting that the isothermal sphere is conformally related to massless global monopole or solid angle deficit spacetime. \\

\section{ Discussion}

Like the uniform density sphere being the prototype of bound fluid distributions representing stellar objects, similarly the isothermal fluid distribution is the prototype of relaxed equilibrium limiting state of cosmological systems -- galaxies \cite{saslaw}, galactic clusters \cite{plummer, zwicky}, or the universe as a whole \cite{sdm}. Even though neither of them is entirely physically satisfactory as the former has sound velocity being infinite while the latter has density diverging at the centre. Despite 
these shortcomings the uniform density sphere and isothermal distributions  seem to capture the essential features of distributions beautifully and that is why all the ballpark parameter values are always based on them. \\

It has been shown \cite{dkm} that
the Schwarzschild interior solution which describes uniform density sphere is also valid for Lovelock gravity of any order $N$ indicating its universality. Since isothermal
distributions enjoy the same status for extended cosmological scale systems, there is a good reason to expect the same be true for them as well; i.e. isothermal sphere solutions  being universal. We have shown here that  indeed  this is the case. The isothermal fluid distribution characterized by the property $p = \alpha \rho \sim 1/r^{2N}$, is described by the same solution for pure Lovelock gravity in all $d\geq 2N+2$ dimensions. It is thus universal for pure Lovelock gravity for a fixed $N$.  \\

Isothermality is characterized by $\rho \sim 1/r^{2N}$, and this property is also shared by spacetime of solid angle deficit \cite{dny} describing the massless  global monopole \cite{bv}. It turns out that the two are indeed conformally related \cite{conform}. Isothermal fluid solution could always be written as conformal to solid angle deficit metric, and conversely a conformal metric to it is shown to describe isothermal distributions for pure Lovelock gravity. We could therefore say that the necessary and sufficient condition for isothermal distribution in pure Lovelock gravity is that it is conformal to solid angle deficit or massless global monopole metric. \\

Since isothermal fluid distribution can occur only in pure Lovelock gravity, it adds a further instance to the long list of phenomena \cite{probes} that establishes uniqueness of pure Lovelock gravity. \\

\bibliography{basename of .bib file}

\end{document}